\documentclass[twocolumn]{aastex62}

\usepackage{graphicx}
\usepackage{dcolumn}
\usepackage{bm}
\usepackage{color}
\newcommand{\msun}{\ensuremath {\rm M}_{\odot}}

\shortauthors{Hoang et al.}

\begin{document}
\title{Detecting Supermassive Black Hole-Induced Binary Eccentricity Oscillations with LISA}

\correspondingauthor{Bao-Minh Hoang}
\email{bmhoang@astro.ucla.edu}

\author{Bao-Minh Hoang}
\affiliation{Department of Physics and Astronomy, University of California, Los Angeles, CA 90095, USA}
\affiliation{Mani L. Bhaumik Institute for Theoretical Physics, Department of Physics and Astronomy, UCLA, Los Angeles,
CA 90095}
\author{Smadar Naoz}
\affiliation{Department of Physics and Astronomy, University of California, Los Angeles, CA 90095, USA}
\affiliation{Mani L. Bhaumik Institute for Theoretical Physics, Department of Physics and Astronomy, UCLA, Los Angeles,
CA 90095}
\author{Bence Kocsis}
\affiliation{Institute of Physics, E\"otv\"os University, P\'azm\'any P. s. 1/A, Budapest, 1117, Hungary}
\author{Will M. Farr}
\affiliation{Department of Physics and Astronomy, Stony Brook University, Stony Brook NY 11794, USA}
\affiliation{Center for Computational Astrophysics, Flatiron Institute, 162 Fifth Avenue, New York, NY 10010, USA}
\author{Jess McIver}
\affiliation{LIGO Laboratory - California Institute of Technology, Pasadena, CA 91125, USA}

\begin{abstract}
Stellar-mass black hole binaries (BHBs) near supermassive black holes (SMBH) in galactic nuclei undergo eccentricity oscillations due to gravitational perturbations from the SMBH. Previous works have shown that this channel can contribute to the overall BHB merger rate detected by the Laser Interferometer Gravitational-Wave Observatory (LIGO) and Virgo Interferometer. Significantly, the SMBH gravitational perturbations on the binary's orbit may produce eccentric BHBs which are expected to be visible using the upcoming Laser Interferometer Space Antenna (LISA) for a large fraction of their lifetime before they merge in the LIGO/Virgo band. For a proof-of-concept, we show that the eccentricity oscillations of these binaries can be detected with LISA for BHBs in the local universe up to a few Mpcs, with observation periods shorter than the mission lifetime, thereby disentangling this merger channel from others. The approach presented here is straightforward to apply to a wide variety of compact object binaries with a tertiary companion.
\end{abstract}

\section{Introduction}
The recent detection of gravitational wave (GW) emission from a merging neutron star binary \citep{GW170817} and merging BHBs \citep{GW150914,GW151226,GW170104,GW170608,GW170814,LIGO+18}
by LIGO/Virgo have ushered in an exciting new era of gravitational wave astrophysics. The astrophysical origin of the detected mergers is currently under debate, with numerous explanations proposed. These explanations can be very roughly divided into two main categories: mergers due to isolated binary evolution \citep[e.g.][]{Belczynski+16,deMink+Mandel,Mandel+deMink,Marchant+16}, and mergers due to dynamical interactions \citep[e.g.][]{Port+00,Wen,OLeary+06,OLeary+09,Oleary+16,Kocsis+Levin,Rodriguez+16,Antonini+Rasio,Askar+17,Fragione+Kocsis18,Wen,AP12,Antonini+14,VanLandingham+16,Hoang+18,Randall+Xianyu,Arca-Sedda+Gualandris,Arca-Sedda+Capuzzo-Dolcetta}.
Orbital eccentricity has been explored as a way to distinguish between these merger channels in both the LIGO/Virgo and LISA frequency bands. In contrast to mergers from isolated binary evolution, merging binaries from dynamical channels have been shown to have measurable eccentricities when they enter the LISA and/or LIGO/Virgo band, and can potentially be used as a way to distinguish between channels \citep[e.g.][]{OLeary+09,Cholis+16,Rodriguez+18,Zevin+18,Lower+18,Samsing18,Gondan+18b,Randall+Xianyu}. Unlike LIGO/Virgo, which can only detect merging BHBs in the final inspiral phase before merger, LISA will be able to detect eccentric stellar-mass BH binaries for long timescales before they merge in the LIGO/Virgo band \citep[e.g.,][]{OLeary+06,Breivik+16,Nishizawa+16,Chen+AmaroSeoane17,Nishizawa+17,Samsing+Dorazio,Dorazio+Samsing,Kremer+18}. This provides us with invaluable insight into the dynamical evolution of eccentric binaries leading up to merger, which has important implications about the astrophysical context in which merging binaries evolve. 

It has been shown that tight binaries orbiting a third body on a much wider ``outer orbit" (a hierarchical triple) can undergo large eccentricity oscillations on timescales longer than the BHB orbital timescale due to gravitational perturbations from the tertiary---the so called eccentric Kozai-Lidov (EKL) mechanism \citep[e.g.,][]{Kozai,Lidov,Naoz16}. In the case of BHBs orbiting a SMBH, high eccentricities can lead to a faster merger via GW emission \citep[e.g.,][]{Wen,Antonini+14,Hoang+18}. Furthermore, this BHB merger channel has been shown to possibly contribute to the overall merger rate at levels comparable to other dynamical channels of mergers \citep{Hoang+18,Hamers+18}. Before they merge, these binaries spend a long time ($10^{2-9}$ yr) oscillating between high and low eccentricities \citep[e.g.,][]{Hoang+18}. 

\begin{figure*}[!t]
\begin{center}
\includegraphics[scale=0.39]{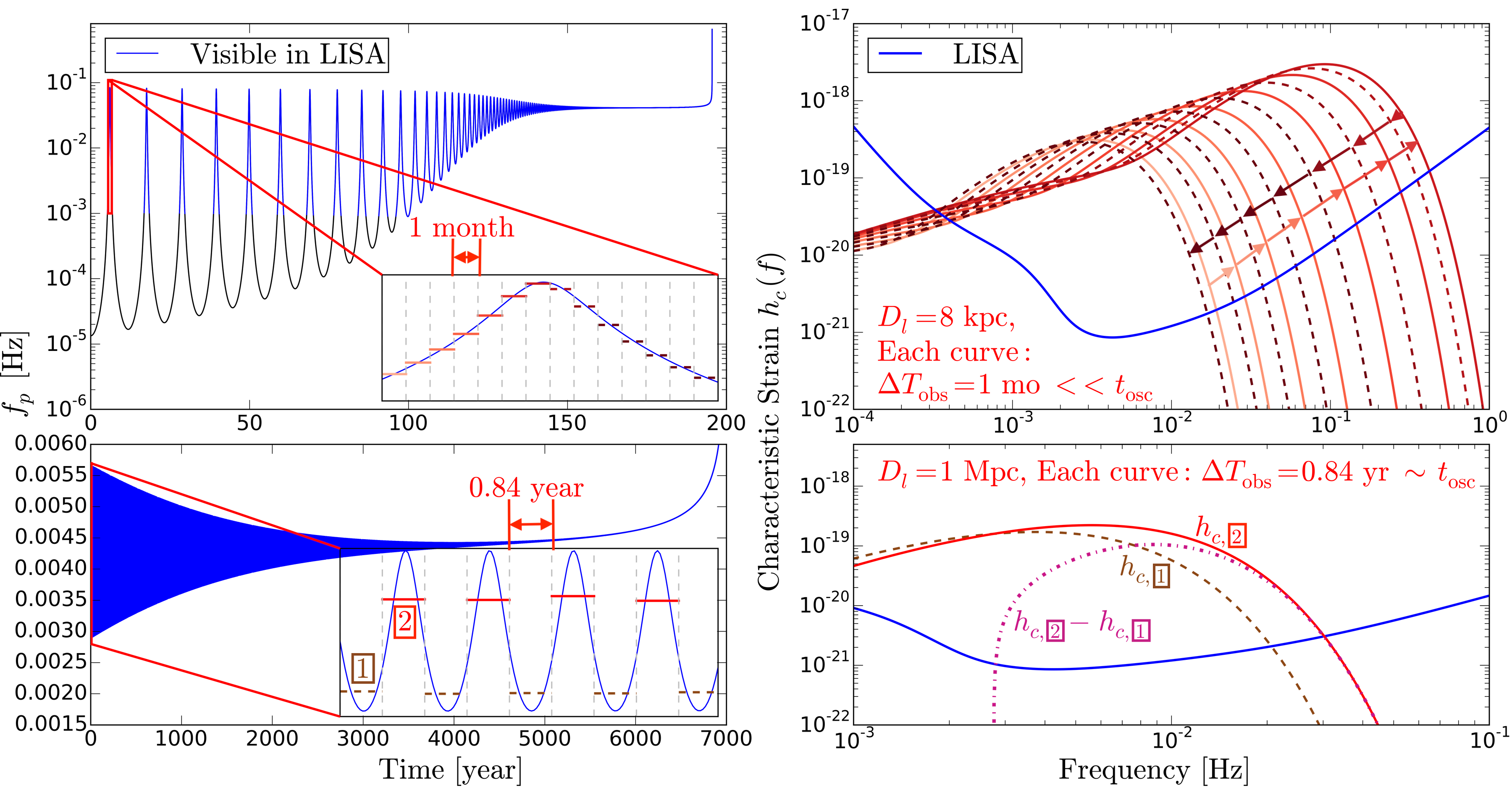}
\qquad\caption{ \upshape \textbf{Two examples of stellar-mass BHBs near a SMBH that exhibits strain oscillations in the LISA band.} {\it Left panels}: the time evolution of the pericenter frequency of two systems undergoing EKL cycles and GW emission. The initial parameters of the top (bottom) binary are $m_1 = 30~\msun$, $m_2 = 20~\msun$, $m_{\bullet} = 4 \times 10^6~\msun$ ($1 \times 10^7~\msun$), $a= 0.15$ ($0.046~\rm AU$), $a_{\rm out} = 250~\rm AU$, $e = 0.5$ ($0.96$), $e_{\rm out} = 0.9$, $i = 88^\circ (90^\circ)$, $g = 0^\circ (135^\circ)$, and $g_{\rm out} = 0^\circ$. We place the top (bottom) system at a distance $D_l$ of 8 kpc (1 Mpc). The proximity of the top system means that it is resolvable with an observation time interval $\Delta T_{\rm obs}$ that is much shorter than the eccentricity oscillation timescale. In contrast, the distance of the bottom system means that a $\Delta T_{\rm obs}$ longer than the eccentricity oscillation timescale is required to resolve the system. This difference of timescales results in two different observables: in the first case the evolution of the average $f_p$ observed per $\Delta T_{\rm obs}$ tracks the evolution of the actual $f_p$ (see inset of top left panel); in the second case, if $\Delta T_{\rm obs}$ is a half integer multiple of the eccentricity oscillation timescale, the average $f_p$ observed per $\Delta T_{\rm obs}$ oscillates (see inset of bottom left panel). {\it Right panels:} Characteristic strain oscillations of the corresponding left panel systems. {\it Top:} First solid strain line has an SNR $\sim 170$. The strain moves to the right as the eccentricity increases (red solid lines), and moves back to the left as the eccentricity decreases due to EKL (red dashed lines). The left panel of the figure shows that this effect can be seen whenever the eccentricity is pumped up to extremal values throughout the binary's life. {\it Bottom:} The brown dashed line $h_c,1$ has an SNR of $\sim 13$ and the observed average $f_p$ oscillates between two values, as shown by the two strain curves $h_{c,1}$ and $h_{c,2}$ (dashed and solid lines, respectively). We have also plotted the difference of the two strains $h_{c,2} - h_{c,1}$ (purple dot-dashed line), which lies well above the LISA sensitivity curve and has an SNR of $\sim 6.4$. 
}
\label{fig:LISAEKL}
\end{center} \vspace{-0.2cm}
\end{figure*}

Eccentric binaries emit GWs over a wide range of frequencies that approximately peaks at a frequency of $f_p(a,e) = (1 + e)^{1/2}(1 - e)^{-3/2} f_{\rm orb}(a)$, with $f_{\rm orb}(a) = (2 \pi)^{-1}\sqrt{G(m_1 + m_2)} a^{-3/2}$, where $m_1$ and $m_2$ are the BHB component masses (we consider mass components of $m_1 = 30~\msun$ and $m_2 = 20~\msun$ for the rest of the paper, which fall well within the mass distribution detected by LIGO/Virgo \citep{LIGO+18}), $a$ is the semi-major axis, $e$ is the orbital eccentricity, and $G$ is the gravitational constant \citep[e.g.][]{OLeary+09}. In the left panels of Figure \ref{fig:LISAEKL} we show the time evolution of $f_p$ for two representative BHB undergoing eccentricity oscillations while orbiting a SMBH of mass $m_\bullet = 4 \times 10^6~\msun$($1 \times 10^7~\msun$), on an outer orbit of semi-major axis $a_{\rm out} = 250~\rm AU$, and eccentricity $e_{\rm out} = 0.9$.  As can be seen from the figure, these BHBs are visible using LISA for a substantial part of their lifetime, providing an unprecedented opportunity for the study of their dynamics. We show in this Letter that the eccentricity oscillations they undergo are detectable with LISA with observational intervals $\Delta T_{\rm obs}$ shorter than the proposed LISA mission lifetime of 4 years \citep{LISAL3}, thereby revealing their astrophysical origin\footnote{We note that while finalizing this manuscript, an independent study by \citet{Randall+Xianyu19} addressed the potential for detecting EKL oscillations with LISA for BHBs in triples. In our proof-of-concept work, we focus specifically on BHBs around an SMBH, and show that these EKL oscillations are indeed significant enough to be detected by LISA. We provide a method that will allow distinguishing a BHB near a tertiary mass, and also between systems with dynamics dominated by GW emission and those that are EKL dominated.}

\section{Two regimes of detectable eccentricity oscillations}
To illustrate the detectability of eccentricity oscillations in GW data, we divide the GW data stream into segments of time duration $\Delta T_{\rm obs}$. 
Generally, to accumulate a sufficient amount of signal-to-noise within individual $\Delta T_{\rm obs}$ intervals, relatively short $\Delta T_{\rm obs}$ 
is sufficient to resolve systems with small $D_l$, where $D_l$ is the luminosity distance of the BHB; conversely, much longer $\Delta T_{\rm obs}$ is required to resolve systems with large $D_l$.  This dichotomy results in two distinct types of observable eccentricity oscillations depending on the ratio of $\Delta T_{\rm obs}$ to the timescale on which the eccentricity oscillates, $t_{\rm osc}$, which depends on both EKL and general relativistic effects \citep[e.g.,][]{Naoz+2013,Randall+Xianyu,Antognini15,Naoz16}. In practice we do not calculate $t_{\rm osc}$, but simply calculate from numerical simulations the value of $\Delta T_{\rm obs}$ which will maximize the detectability of eccentricity oscillations (details are given later in this Letter).

Here we focus on two cases:
\begin{enumerate}
\item $\Delta T_{\rm obs} \ll t_{\rm osc}$: For BHBs at small luminosity distances $D_l$, the signal-to-noise accumulates rapidly and we can divide the data stream into short $\Delta T_{\rm obs}$ intervals that are much smaller than $t_{\rm osc}$ and measure the eccentricity separately for each interval. If the change in eccentricity between successive intervals is larger than the eccentricity measurement accuracy in each interval, the evolution of eccentricity can be detected by comparing the eccentricities in different observation intervals. We show an example of this in the top panels of Figure \ref{fig:LISAEKL}, with a BHB 8 kpc away, with the data stream divided into intervals of $\Delta T_{\rm obs} = 1$ month.
\item $\Delta T_{\rm obs} \sim t_{\rm osc}$ or $\Delta T_{\rm obs} > t_{\rm osc}$: For BHBs at greater $D_l$ with weaker GWs, 
the data stream must be processed in longer $\Delta T_{\rm obs}$ intervals, that may be comparable to or larger than $t_{\rm osc}$. In this case, the eccentricity evolution cannot be measured as accurately as in the $\Delta T_{\rm obs} \ll t_{\rm osc}$ case. Nevertheless, the eccentricity variation between different $\Delta T_{\rm obs}$ intervals may still cause a significant change in the GW waveform that can be measured. In the bottom panels of Figure \ref{fig:LISAEKL} we show eccentricity oscillations in a BHB 1 Mpc away, with the data stream divided into intervals of $\Delta T_{\rm obs} = 0.84$ year ($\sim 1/2$ times the BHB's $t_{\rm osc}$). The change in the GW signal is well above the LISA noise.
\end{enumerate}

Below we quantify the parameter space where stellar-mass BHBs are resolvable with LISA, and where their eccentricity oscillations are large enough to be detectable with LISA.

\begin{figure*}
\includegraphics[width=\linewidth]{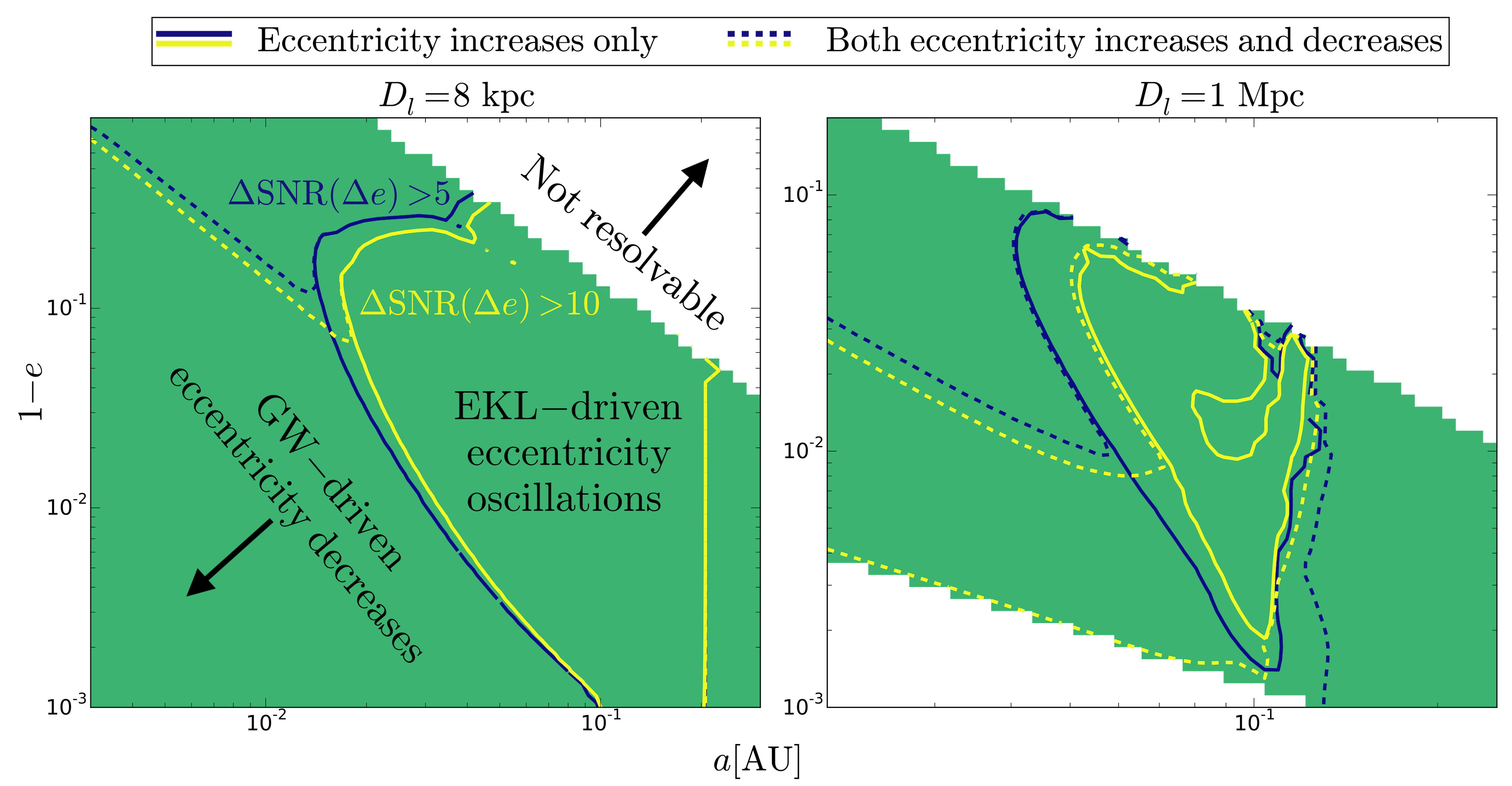}
\caption{{\bf Map of the parameter space where SMBH-induced eccentricity oscillations are visible with LISA.} In the left (right) panel we show the case for $D_l = 8$ kpc ($D_l = 1$ Mpc). The green area shows where systems are resolvable with ${\rm SNR} \geq 5$ for $\Delta T_{\rm obs} \leq 5$ years. Each BHB is placed on an orbit of $a_{\rm out} = 250$ AU and $e_{\rm out} = 0.9$ around a SMBH of mass $m_{\bullet} = 4 \times 10^6 (1 \times 10^7)~\msun$ for the $D_l = 8$ kpc ($D_l = 1$ Mpc) case. The blue (yellow) contours enclose the areas where eccentricity changes are detectable at the level of $\Delta {\rm SNR}(\Delta e) > 5 (10)$. The solid contours enclose only the parameter space where the eccentricity {\it increases} between consecutive $\Delta T_{\rm obs}$ intervals (due to SMBH-induced EKL) are detectable. The dashed contours enclose the parameter space where both eccentricity {\it increases} and {\it decreases} are detectable. The eccentricity decrease can either be caused by GW emission or by EKL oscillations, which occupy two distinct regions of the parameter space (see plot labels). We note that going beyond a distance of a few Mpcs reduces the EKL-driven area to a negligible part of the parameter space, and is omitted here to avoid clutter. The sharp cutoff to the right of the contours on the left panel comes from taking into account the Hill stability criterion.}
\label{fig:SNRgrid}
\end{figure*}

\section{Detectability of eccentric stellar-mass BHBs with LISA}
Unlike circular binaries, which emit GWs at a single frequency, equal to twice the orbital frequency, eccentric binaries emit at a wide range of orbital frequency harmonics. The complex GW dimensionless strain $h(a,e,t)$  of a binary of semi-major axis $a$ and eccentricity $e$ is then the sum of the strains at each orbital frequency harmonic $f_n = n f_{\rm orb}$ \citep{Peters+63}. We follow the calculation of the GW strain from \citep{Kocsis+12}, where $h(a,e,t)$ is defined as:
\begin{equation}\label{eq:straintime}
h(a,e,t) = \sum_{n = 1}^{\infty} h_n(a,e,f_n){\rm exp}(2 \pi i f_n t) \ ,
\end{equation}
where 
\begin{equation}
h_n(a,e,f_n) = \frac{2}{n}\sqrt{g(n,e)}h_0(a)\ ,
\end{equation}
with $h_0(a)$ representing the dimensionless strain amplitude for circular binary orbit with masses $m_1$ and $m_2$ at a luminosity distance of $D_l$, averaged over the binary orientation, i.e., 
\begin{equation}
h_0(a) = \sqrt{\frac{32}{5}}\frac{G^2}{c^4}\frac{m_1 m_2}{D_l a} \ ,
\end{equation}
where $c$ is the speed of light and $g(n,e)$ is defined as: 
\begin{eqnarray}
g(n,e) &= & \frac{n^4}{32}\Big[( J_{n-2} - 2 e J_{n - 1} + \frac{2}{n} J_n + 2 e J_{n+1} - J_{n+2})^2\nonumber \\ &+& (1 - e^2)(J_{n-2} - 2 J_n + J_{n+2})^2 + \frac{4}{3 n^2}J_n^2\Big],
\end{eqnarray}
where $J_i$ is the $i$th Bessel function evaluated at $ne$ \citep{Peters+63}. We have neglected a factor of $(1 + z)^2$ in $h_0(a)$ that accounts for the effects of Doppler shift due to the peculiar velocity of the source and cosmological redshift. The effects of peculiar velocity and redshift are equivalent to a change of apparent distance and object masses \citep{Kocsis+06}. However, since the furthest luminosity distances considered in this paper are a few Mpcs, the corresponding redshift $z$ is very small and we do not expect this effect to significantly alter our results. 

We may crudely approximate the characteristic strain of an evolving eccentric binary using the Fourier transform of a stationary binary as:

\begin{equation}\label{eq:hc}
h^2_c(a,e,f) = 4 f^2 |\tilde{h}(a,e,f)|^2\times {\rm min}\Big(1,\frac{f_n}{\dot{f}_n}\frac{1}{\Delta T_{\rm obs}}\Big),
\end{equation}
 
where $\tilde{h}(a,e,f)$ is the Fourier transform of Equation (\ref{eq:straintime}) over an observational period of $\Delta T_{\rm obs}$, and the factor inside the minimum function accounts for the fact that the signal power actually only accumulates for a time of $\min(\Delta T_{\rm obs}, f_n/\dot{f}_n)$ in each frequency bin as $a$ and $e$ vary slowly \citep{Cutler+94,Flanagan+98}. We show $h_c(f)$ in the right panels of Figure \ref{fig:LISAEKL} for the corresponding systems in the left panels. As can be seen from these figures, the strain spectrum will visibly oscillate with different $f_p$ peak frequency due to the underlying eccentricity oscillations.

We note that a hallmark feature of EKL is oscillations in $i$---the inclination of the binary angular momentum vector with respect to the angular momentum vector of the outer orbit---that are out of phase with oscillations in $e$ \citep[e.g.][]{Naoz16}. Furthermore, as LISA orbits around the Sun, the angle between the BHB angular momentum and the line of the sight will also change. These combined effects result in variations in the binary inclination with respect to the line of sight, which we have neglected in the calculation above. However, variations in binary inclination will only modulate the amplitude of the signal without changing $f_p$. Thus, changes in $f_p$ due to changes in $e$ will still be detectable. Furthermore, the signal amplitude modulations due to oscillations in $i$ may themselves be used to indicate the presence of EKL. Another effect that we have neglected in our strain calculation is the precession of the BHB pericenter due to both EKL and general relativity. Pericenter precession, like inclination oscillations, does not effect $f_p$, so we do not expect it to significantly alter the conclusions of this Letter. However, it will change the polarization of the waveform, which may also independently indicate the presence of EKL. We leave these considerations to a future study.

To quantify the parameter space where these binaries are detectable in LISA, we compute the signal-to-noise ratio (SNR) as a function of $a$ and $e$ as: \citep[e.g.][]{Robson+18}:
\begin{equation}\label{eq:SNR}
{{\rm SNR}^2(a,e)} = \int \frac{h^2_c(a,e,f)}{f^2 S_n(f)} df \ , 
\end{equation}
where $S_n(f)$ is the effective noise power spectral density of the detector, weighted by the sky and polarization-averaged signal response function of the instrument  \citep[e.g., equation 1 in][]{Robson+18}. In the case that the LISA mission lifetime is extended to 10 years \citep[e.g.,][]{LISAL3}, and assuming that at least two $\Delta T_{\rm obs}$ intervals is required to detect a change in eccentricity, we pick a maximum possible $\Delta T_{\rm obs}$ of 5 years. Furthermore, we set the threshold for resolvability at SNR = 5.  In Figure \ref{fig:SNRgrid} we show in green the region in $a$ and $1 - e$ parameter space of the inner binary where ${\rm SNR} \geq 5$ is achievable for $\Delta T_{\rm obs} \leq 5$ years, for $D_l = 8$ kpc and $D_l = 1$ Mpc, respectively. In this initial SNR calculation we have neglected eccentricity evolution due to EKL.

\section{Detectability of eccentricity evolution}
Having established the parameter space where eccentric stellar-mass BHBs are visible to LISA, we now quantify LISA's ability to detect eccentricity changes in these binaries. We do this by finding the parameter space where the change in eccentricity, $\Delta e$, induces a change in the waveform that has a signal-to-noise $\Delta {\rm SNR}(\Delta e)$, of 5 or greater. 

We run simulations of BHBs orbiting a SMBH, with $a$ and $e$ sampled from the parameter space shown in Figure \ref{fig:SNRgrid}. In our simulations we include the secular equations up to the octupole level of approximation \citep[e.g.,][]{Naoz+11sec}, general relativity precession of the inner and outer orbits \citep[e.g.,][]{Naoz+2013}, and GW emission \citep{Peters64}. We consider a SMBH of mass $4 \times 10^6 (1 \times 10^7)~\msun$ for the $D_l = 8$ kpc ($D_l = 1$ Mpc) case, and nominal outer orbit parameters $a_{\rm out} = 250$ AU, $e_{\rm out}$ = 0.9, and $\omega_{\rm out} = 0^\circ$, where $\omega_{\rm out}$ is the argument of pericenter of the outer orbit. For the most comprehensive estimate of the region where $\Delta e$ is large enough to result in a significant $\Delta {\rm SNR}(\Delta e)$, we choose the mutual inclination $i$ to be $90^\circ$, which maximizes the amplitude of eccentricity oscillations in a majority of cases. All that remains is to choose $\omega$, the argument of pericenter of the inner orbit, which sets the phase of the oscillation and partially determines whether the change in eccentricity between consecutive $\Delta T_{\rm obs}$ is detectable. We run our simulations for three different values of $\omega$: $0^\circ{}$, $45^\circ{}$, and $135^\circ{}$ \citep[e.g.,][]{Li+14Chaos}. Lastly, we have restricted $a$ in our simulations to be less than ${\min}[a_h,a_R]$, where $a_h = 0.1~a_{\rm out} (1 - e^2_{\rm out})/e_{\rm out}$, and $a_R = a_{\rm out} ((m_1 + m_2)/3/m_{\bullet})^{1/3}(1 - e_{\rm out})/(1 + e)$. The first is the condition to be a hierarchical triple, so that our secular equations of motion are applicable \citep[e.g.,][]{Naoz16}, and the second is the condition that the BHB does not cross the Roche limit of the SMBH \citep[e.g.,][]{NaozSilk}. In all cases shown in Figure \ref{fig:SNRgrid} we find that $a_R < a_h$. This condition causes the sharp cutoff in the right side of the contours seen in the left panel of Figure \ref{fig:SNRgrid}). Systems near this line may develop Hill instabilities as the inner BHB's eccentricity is excited, and be shorter lived than assumed here.

We then search through these simulations to find the value of $\Delta T_{\rm obs}$ that will maximize $\Delta {\rm SNR}(\Delta e)$. We restrict $\Delta T_{\rm obs}$ to be greater than the value of $\Delta T_{\rm obs}$ that will give an SNR of 5, and to be smaller than 5 years. We also maximize $\Delta {\rm SNR}(\Delta e)$ with respect to $\omega$, although we note that, in this proof-of-concept calculation, we have only sampled three fixed values of $\omega$. Thus, we expect that the estimate given here is an underestimate of the parameter space where eccentricity oscillations are detectable. $\Delta {\rm SNR}(\Delta e)$ is calculated by time-averaging the eccentricity evolution $e(t)$ over each $\Delta T_{\rm obs}$ interval, and using these averaged eccentricities to calculate the change in SNR between the two intervals using Equation (\ref{eq:SNR})\footnote{Note that for simplicity we have assumed a constant $a$ over consecutive $\Delta T_{\rm obs}$ intervals. This approximation holds well for most of the EKL-driven systems. However, for systems for which GW-emission is significant, $a$ is shrinking over $\Delta T_{\rm obs}$ (for example, far left systems in Figure \ref{fig:SNRgrid})}. In Figure \ref{fig:SNRgrid} we  show contours of where $\Delta {\rm SNR}(\Delta e)$ is greater than $5$ and $10$. We  distinguish between the cases where $\Delta {\rm SNR}(\Delta e)$ results from an {\rm increase} in eccentricity, and where $\Delta {\rm SNR}(\Delta e)$ results from a {\rm decrease} in eccentricity. Whereas the former can only be caused by EKL in our simulations, the latter can be caused by either EKL or GW-emission. However, the two different types of eccentricity decreases occupy very distinct parts of the parameter space, as shown in \ref{fig:SNRgrid}. We can see that EKL-driven eccentricity oscillations are detectable for a large fraction of the BHB parameter space, out to a few Mpcs.

\section{Discussion}
We present here a novel approach to distinguish eccentric stellar mass BHBs that undergo eccentricity oscillations induced by a SMBH from other sources of GWs. It has been suggested that stellar binaries exist in high abundance in the vicinity of our galactic center, and thus also in other galactic nuclei \citep{Ott+99,Martins+06,Pfuhl+14,Stephan+16,Naoz+18,Hailey+18,Stephan+19}.  In particular, \citet{Stephan+19} showed that the formation rate of compact object binaries (including EKL) is about $10^{-6}$~yr$^{-1}$ at the center  of a Milky-Way--like galaxy. Assuming a galaxy density of Milky-Way like galaxies is $0.02$~$~$Mpc$^{-3}$ \citep{Conselice+2005}, we find that inside the local group sphere ($\sim 3$~Mpc, where we expect eccentricity oscillations to be detectable\footnote{Note that eccentric binaries themselves can be detected in LISA to much larger distances \citep[e.g.,][]{Fang+19}.}), the rate of formation of compact object binaries is $\sim2\times 10^{-6}$~yr$^{-1}$. EKL contributes to the merger, and therefore depletion, of BH binaries after about $10^8$~yr \citep[e.g,][]{Hoang+18}. Thus, if we assume that all binaries in galactic nuclei are depleted due to EKL, we estimate that about $200$ binaries may be in the relevant parameter space detectable by LISA. On the other hand, if we assume that no binaries are depleted due to EKL, we have that over the lifetime of the local group ($\sim 10$~Gyr), there are potentially  $\sim 20,000$ binaries that can have their eccentricity evolution detected in LISA. The true number is likely between these two limits.

In this proof-of-concept calculation, we have shown that eccentricity changes in a stellar-mass BHB induced by gravitational perturbations from a nearby SMBH is detectable by LISA (e.g., Figures \ref{fig:LISAEKL} and \ref{fig:SNRgrid}). This could be used as a method of distinguishing these GW sources from sources in other astrophysical contexts. Constraining the binary's eccentricity and semi-major axis with LISA's future waveform templates could disentangle the evolutionary path of the system. Notably, detecting a binary in the EKL-driven regime can infer the existence of a nearby SMBH (or another tertiary). Furthermore, some of the physical parameters of the system, such as tertiary mass, eccentricity, and semi-major axis can be constrained.

It is not unlikely that the LISA mission lifetime will be extended beyond 4 years. A $10$~year life time can potentially broaden the region in parameter space where eccentricity oscillations are detectable, as well as the distance to which they are detectable. However, we note that in the $D_l = 8$ kpc case, these eccentricity oscillations can be detected on a timescale of months (see upper two panels of Figure \ref{fig:LISAEKL}). We found that the parameter space in which EKL-driven oscillations can be detected extends to a distance of a few Mpcs.

We note that for this proof-of-concept calculation we adopted the secular approximation, however, some of the systems in Figures 1 and 2 deviate from pure secular dynamics. These systems may exhibit rapid (orbital time scales) eccentricity oscillations \citep[e.g.][]{Ivanov+05,Antognini+13,Antonini+14}, which are at very low amplitude compared to the envelope eccentricity oscillations and may average out. We leave it to a future study to investigate whether these non-secular oscillations are significant. Furthermore, the deviation of the secular approximation may ultimately result in higher eccentricity spikes than the one calculated using the secular approximation \citep[e.g.][]{Katz2011,Bode+13}, which will only strengthen the overall effect, but potentially complicate the analysis. We estimate the region of parameter space where non-secular effects might become important as where the EKL oscillation timescale is within a factor of a few larger than the outer orbital period (for a definition of the EKL timescale, see \citet{Naoz16}). For a SMBH of mass $4 \times 10^6~\msun$ and the nominal orbital parameters assumed in Figure \ref{fig:SNRgrid}, we find that the EKL oscillation timescale will be within a factor of 2 (5) of the outer orbital period for $a_{\rm out} \gtrsim 0.16~(0.12)$ AU. As can be seen in Figure \ref{fig:SNRgrid}, these non-secular effects, if at all detectable, will be most likely detected at small $D_l$ like $8$ kpc. Finally, we note that some of these systems may be shorter lived than assumed here since as $e$ increases the binary may cross the SMBH Hill radius (as noted in other systems,\citep[e.g.][]{Li+15}).

The methodology presented here is straightforward to extend to stellar-mass BHBs around any tertiary, BH-SMBH binary with an SMBH companion, as well as triples containing any compact objects such as ones containing white dwarfs and neutron stars. The EKL mechanism is pervasive for a large set of astrophysical scenarios \citep[e.g.][]{Ford00,Naoz16} and for a wide range of triple masses. Thus, the detection of EKL in different triple systems with LISA may allow us to distinguish between different triple orbital configurations, in particular, the tertiary's mass, outer orbit separation, eccentricity, and inclination. Localization within a galaxy will further allow disentanglement between the orbital parameters. Additionally, the approach shown here can help disentangle between binaries in triples and binaries of non-triple origin, since the latter will not exhibit oscillations in the characteristic strain-frequency parameter space. Thus, the proposed methodology here can serve as a potentially powerful method to disentangle different GW sources.

Furthermore, in this proof-of-concept Letter we have only focused on one effect of EKL---the oscillation in eccentricity---when there are in fact other EKL-induced effects like oscillation in inclination and precession of pericenter that can leave detectable imprints on the GW waveform. Thus, the detectability of EKL with GW may be possible for a wider range of systems than predicted by this work.

\section*{Acknowledgements}
B.M.H and S.N. acknowledge the partial support of NASA grant No.--80NSSC19K0321. S.N. also thanks Howard and Astrid Preston for their generous support. This project has received funding from the European Research Council (ERC) under the European Union's Horizon 2020 research and innovation programme under grant agreement No 638435 (GalNUC) and by the Hungarian National Research, Development, and Innovation Office grant NKFIH KH-125675 (to B.K.).

\bibliography{Binary}
\end{document}